\documentclass[aps,prb,twocolumn,amsmath,amssymb,superscriptaddress]{revtex4}
\usepackage{graphicx,dcolumn,bm,epstopdf,comment,color,multirow}

\def \SFOO {Sr$_2$FeOsO$_6$}
\def \SIOO {Sr$_2$InOsO$_6$}
\def \SYOO {Sr$_2$YOsO$_6$}
\def \SCOO {Sr$_2$CoOsO$_6$}

\begin{document}
\title{Muon-spin relaxation study of the double perovskite insulators Sr$_{2}B$OsO$_6$ ($B={\rm Fe, Y, In}$) }
 \author{R. C. Williams}
 \email{r.c.williams@durham.ac.uk}
  \author{F. Xiao}
  \author{I. O. Thomas}
  \author{S. J. Clark}
  \author{T. Lancaster}
\affiliation{Durham University, Department of Physics, South Road,
  Durham, DH1 3LE, UK}
\author{G. A. Cornish}
\affiliation{Department of Physics, University of York, Heslington, York, YO10 5DD, UK}  
\author{S. J. Blundell}
\author{W. Hayes}
\affiliation{Oxford University Department of Physics, Parks Road, Oxford, OX1 3PU, UK}
\author{A. K. Paul}
\altaffiliation{Present address: National Institute of Technology, Kurukshetra, 136119, India }
\affiliation{Max-Planck-Institut f\"{u}r Chemische Physik fester Stoffe, N\"{o}thnitzer Str. 40, D-0187 Dresden, Germany}
\affiliation{Max-Planck-Institut f\"{u}r Festk\"{o}rperforschung,
  Heisenbergstr.\ 1, D-70569 Stuttgart, Germany}
\author{C. Felser}
\affiliation{Max-Planck-Institut f\"{u}r Chemische Physik fester Stoffe, N\"{o}thnitzer Str. 40, D-0187 Dresden, Germany}
\author{M. Jansen}
\affiliation{Max-Planck-Institut f\"{u}r Chemische Physik fester Stoffe, N\"{o}thnitzer Str. 40, D-0187 Dresden, Germany}
\affiliation{Max-Planck-Institut f\"{u}r Festk\"{o}rperforschung,
  Heisenbergstr.\ 1, D-70569 Stuttgart, Germany}  

\date{\today}

\begin{abstract}
We present the results of zero-field muon-spin relaxation measurements made on the double perovskite insulators Sr$_{2}B$OsO$_6$ ($B={\rm Fe, Y, In}$). Spontaneous muon-spin precession indicative of quasistatic long range magnetic ordering is observed in Sr$_{2}$FeOsO$_6$ within the AF1 antiferromagnetic phase for temperatures below $T_{\rm N}=135 \pm 2~{\rm K}$. Upon cooling below $T_2 \approx 67~{\rm K}$ the oscillations cease to be resolvable owing to the coexistence of the AF1 and AF2 phases, which leads to a broader range of internal magnetic fields. Using density functional calculations we identify a candidate muon stopping site within the unit cell, which dipole field simulations show to be consistent with the proposed magnetic structure. The possibility of incommensurate magnetic ordering is discussed for temperatures below $T_{\rm N}=53~{\rm K}$ and 25~K for Sr$_{2}$YOsO$_6$ and Sr$_{2}$InOsO$_6$, respectively.

\end{abstract}
\maketitle

\section{Introduction}

Compounds exhibiting a perovskite structure are among the most intensively studied materials in condensed matter physics. The $AB$O$_3$ structure comprises corner-sharing $B$O$_6$ octahedra
and a large alkali, alkali earth metal or rare earth metal cation $A$. The complex interplay between spin, orbital, charge and lattice degrees of freedom leads to a rich variety of physical phenomena including multiferroicity [\onlinecite{tokunaga09,ramesh07,cheong07}], superconductivity [\onlinecite{cava88}] and topological insulators [\onlinecite{jin13}].

Further scope for tunability and the engineering of magnetic properties is provided by partial cation substitution, including the realisation of the ordered double perovskite (DP) structure $A_{2}BB'$O$_6$, with interpenetrating quasi-face-centred-cubic (fcc) sublattices [\onlinecite{vasala14}]. Choosing a combination of $3d$ and $5d$ transition metals (TMs) for the cations $B$ and $B'$ makes a plethora of exotic physical properties possible: for example the half metals Sr$_2$FeMoO$_6$ and Sr$_2$FeReO$_6$ which display colossal magnetoresistance [\onlinecite{kobayashi98}] and room-temperature tunneling magnetoresistance [\onlinecite{kobayashi99}], respectively.
The relative occupancy, symmetry and energy scales of the $d$-orbitals on the $B$ and $B'$ sites determines the nature of the magnetic interactions, through a competition between on-site Coulomb repulsion (described by the Hubbard $U$ and Hund's rules), which would favour a superexchange mechanism between localised spins, and kinetic energy reduction via electron delocalisation which favours double exchange (DE). The impact of the large spatial extent of 5$d$ orbitals is further complicated by the role of spin-orbit (SO) interactions, which have been shown to produce interesting states including an unconventional Mott insulator phase in TM oxides [\onlinecite{kim08,kim09}]. Taken together, these complications mean that a straightforward application of the Goodenough Kanamori Anderson (GKA) rules for superexchange is not possible.

\begin{figure*}[ht] 	
\centering
\includegraphics[width=0.75\linewidth]{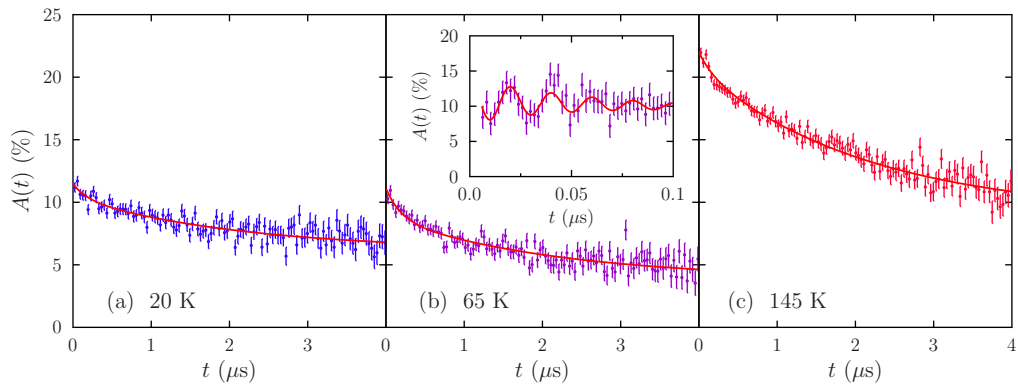}
\caption{ \SFOO{} ZF $\mu^+$SR asymmetry data measured for (a) $T= 20~{\rm K}$, (b) 65~K and (c) 145~K. Inset to (b): typical oscillations visible at early times in the intermediate temperature region. Solid lines are fits described in the main text. \label{SFOOraw}}
\end{figure*}

Recently, muon-spin relaxation ($\mu$SR) was employed to help characterise the exotic magnetism present in the insulating osmate DP \SCOO{}, where site-specific magnetic dynamics lead to independent magnetic ordering on the two sublattices [\onlinecite{yan14,morrow13,paul13c}]. This has motivated our present $\mu$SR study into the related compound \SFOO{} (SFOO hereafter), which exhibits some intriguing magnetic phenomena. Physical and magnetic characterisation of SFOO shows that it is a narrow gap semiconductor where the crystal assumes the tetragonal $I4/m$ space group [\onlinecite{paul13a}], as predicted by first principles calculations [\onlinecite{wang11}]. The system assumes an antiferromagnetically ordered (AFM) state (labeled AF1) below the N\'{e}el temperature $T_{\rm N}=140~{\rm K}$, which precedes a magneto-structural transition upon further cooling at $T_2=67~{\rm K}$ (labeled AF2) [\onlinecite{adler13}]. The ordered magnetic spin structure in this low-$T$ phase adopts a different sequence of spin orientations along the tetragonal $c$-axis due to a lattice instability which supports a bifurcation of iron-osmium distances, reflecting the subtle balance between degrees of freedom and competing exchange interactions in this system [\onlinecite{paul13b}]. Detailed first principles studies revealed the importance of AFM out of plane next-nearest neighbour (nnn) interactions between the extended Os $5d$ orbitals, which compete with ferromagnetic (FM) nearest neighbour (nn) Fe-Os DE interactions and introduce strong frustration [\onlinecite{kanungo14}]. This is believed to be the dominant factor in driving the system to the AF2 spin structure, where spins along the crystallographic $c$-axis order in an alternating $\uparrow \uparrow \downarrow \downarrow$ sequence.

The fragility of the FM DE interaction in SFOO has been demonstrated using the application of external pressure, where compression leads to a reversal in sign of the Fe-Os exchange interaction along the $c$-axis as the AFM superexchange interaction between $t_{2g}$ orbitals overcomes the FM DE interaction between $e_g$ orbitals to induce an overall ferrimagnetic ground state, with considerable remanent magnetisation and coercivity [\onlinecite{veiga15}].

In the systems \SYOO{} and \SIOO{} (hereafter SYOO and SIOO, respectively) $B$ is a non-magnetic $4d$ TM ion ($4d^0$ and $4d^{10}$ for Y$^{3+}$ and In$^{3+}$, respectively), leaving just a fcc lattice of magnetic Os$^{5+}$ ($5d^{3}$) ions [\onlinecite{paul15}]. AFM nn interactions ($J_{\rm nn}$) within this structure are expected to create strong geometric frustration which precludes long range magnetic ordering (LRO). However, the degeneracy of ground state spin structures may be alleviated through either long ranged nnn interactions ($J_{\rm nnn}$) or magnetic anisotropy [\onlinecite{kuzmin03}], leading to magnetic ordering below suppressed onset temperatures of $T_{\rm N}=53~{\rm K}$ and 26~K for SYOO and SIOO, respectively [\onlinecite{paul15}]. The presence of frustration may be inferred from the large ratio between the Weiss constant and ordering temperature $| \theta | / T_{\rm N}$ for these compounds.

In this paper we present the results of zero-field $\mu$SR measurements made on the three compounds: \SFOO{}, \SYOO{} and \SIOO{}. The structure of the paper is as follows: In section \ref{sec:exp} we explain the experimental procedure. In section \ref{sec:results} we present the $\mu$SR results. In section \ref{sec:site} we use density functional theory and dipole field simulations to identify candidate muon stopping sites and discuss the possibility of incommensurate ordering in SYOO and SIOO. Finally, in section \ref{sec:conc} we present our conclusions.

\section{Experimental}
\label{sec:exp}

In a $\mu$SR experiment [\onlinecite{blundell99}] spin-polarised positive muons are implanted in a sample and subsequently decay into a positron with average lifetime $\tau \approx 2.2 ~\mu \mathrm{s}$. The decay positron is emitted preferentially in the direction of the muon's instantaneous spin direction, and the measured quantity of positron asymmetry $A(t)$ is proportional to the muon ensemble's spin polarisation, the time dependence of which is sensitive to the local magnetic field experienced at the muon stopping site. Zero-field $\mu$SR measurements of the double perovskite compounds SFOO, SYOO and SIOO were carried out on powder samples using the GPS spectrometer at S$\mu$S, Paul Scherrer Institut, Switzerland. The samples were packed in Ag envelopes (foil thickness 25~$\mu$m) and taped to a silver fork in the so-called `fly-past' configuration. 

\section{Results}
\label{sec:results}

\section*{\SFOO{}}

\begin{figure}[t]
\centering
\includegraphics[width=\linewidth]{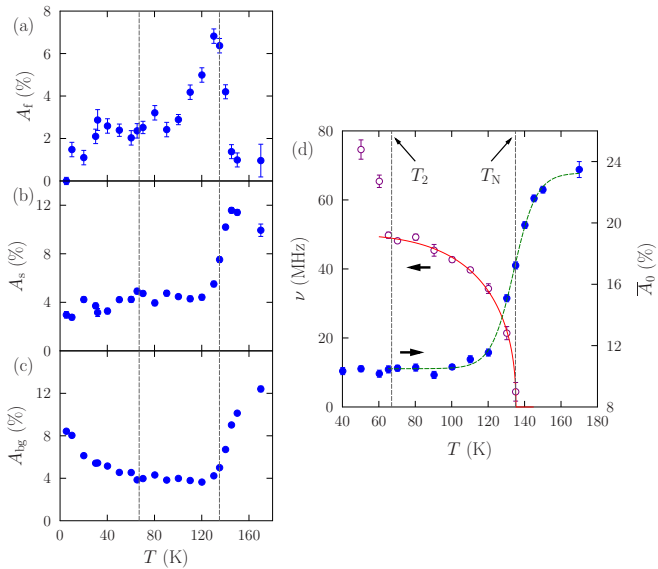}
\caption{\label{fig:SFOOfit} Fitted values of (a) fast relaxation, (b) slow relaxation and (c) baseline component amplitudes from equation~\ref{2exps}, for \SFOO{}. The dashed grey lines indicate the N\'{e}el temperature $T_\mathrm{N}=135~\mathrm{K}$ and the secondary transition temperature $T_2=67~\mathrm{K}$. (d) Fitted precession frequencies obtained using equation~\ref{eqn:osc} and early-time average asymmetry. Solid and dashed lines are fits to the phenomenological expression equation~\ref{eqn:freq}, and the broadened step function equation~\ref{eqn:FD}, respectively. }
\end{figure}

Example asymmetry data for SFOO are shown in figure~\ref{SFOOraw}. Oscillations in the measured asymmetry are resolvable for early times ($t \lesssim 0.1~{\rm \mu s}$) in the temperature range $60\lesssim T \lesssim 135~{\rm K}$ [see figure~\ref{SFOOraw}(b) inset], which is unambiguous evidence for LRO of magnetic moments. Asymmetry averaged over early-times, i.e.\  $\overline{A}_0 \equiv \langle A (t \leq 0.1~{\rm \mu s}) \rangle$, is shown in figure~\ref{fig:SFOOfit}(d). For low temperatures the initial asymmetry is approximately constant. However, around $120~\mathrm{K}$ it increases dramatically upon warming across the transition region around $T_{\mathrm{N}}$. This recovery of initial asymmetry reflects the gradual destruction of LRO.  There are no marked features in $\overline{A}_0$ around the secondary transition at $T_2$. The broadened step-function
\begin{equation} \label{eqn:FD}
\overline{A}_0(T) = A_2 + \frac{A_1  -A_2}{{\rm e}^{(T-T_{\rm mid}) / w}}
\end{equation}
may be used to parametrise the smooth transition between high- (low-) temperature asymmetry values $A_2$ ($A_1$) with midpoint $T_{\rm mid}$ and width $w$ [\onlinecite{steele11}]. The resultant fit for $T \geq 65~{\rm K}$ data sets yielded $T_{\rm N} = T_{\rm mid} \pm w = 134 \pm 6~{\rm K}$.

Below $T_{\mathrm{N}}$ the presence of large quasistatic, or slowly fluctuating, magnetic fields at the muon stopping sites (with components perpendicular to the muon polarisation direction) leads to a rapid dephasing of spin coherence. In the fast fluctuation limit, relaxation rates vary as  $\lambda \propto \langle B^2 \rangle \tau$, where $1/\tau$ is representative of a fluctuation rate associated with dynamics [\onlinecite{hayano79}], and the second moment of the field distribution $\langle B^2 \rangle$ is expected to be large for an ordered magnetic system comprising two magnetic species. Components with relaxation rates of several hundred $\mu{\rm s}^{-1}$ will not be resolved and so are `lost' from the asymmetry data [figures \ref{SFOOraw}(b) and (c)]. The component of asymmetry which is lost upon cooling through $T_{\rm N}$ is $A_{\rm lost} = A_2 - A_1 = 12.8 \pm 0.4 \%$.

In order to parametrise the behaviour across all available temperatures, data for all times ($t\leq 9.7~{\rm \mu s}$) were heavily binned (such that the histogram interval width was $\Delta t = 14.6~{\rm ns}$) and fitted to the relaxation function
\begin{equation} \label{2exps}
A(t) = A_{\rm s} e^{-\lambda_{\rm s} t} + A_{\rm f} e^{-\lambda_{\rm f} t} + A_{\rm bg}.
\end{equation}
Here, s and f denote slow (small $\lambda$) and fast (large $\lambda$) relaxation components, respectively, and $A_{\rm bg}$ is a time-independent baseline contribution arising from muons which stop in the silver sample holder, or whose spin lies parallel to a static magnetic field and therefore do not depolarise. We note that $A_{\rm bg}$ may exhibit a small systematic temperature dependence due to the thermal expansions of cryostat and sample holder components, however these effects are not expected to be significant for this frequently used experimental arrangement. The fitted relaxation rates $\lambda_{\rm s,f}$ were found to be approximately constant across the temperature range, to within uncertainties, and so were fixed to their average values of $0.46~{\rm MHz}$ and $5.4~{\rm MHz}$, respectively.

The fitted values of the component amplitudes are displayed in figures~\ref{fig:SFOOfit}(a)--(c). All three components undergo large changes of amplitude around $T_{\rm N} \approx 140~{\rm K}$ consistent with magnetic ordering. The width of the transition region (approximately $10~{\rm K}$) indicates a degree of static or dynamic disorder in the ordered moments. The baseline and slow relaxation amplitudes $A_{\rm bg}$ and $A_{\rm s}$ both sharply decrease from their high temperature values at around $T_{\rm N}$. Whilst $A_{\rm s}$ does not undergo any further changes, the baseline amplitude $A_{\rm bg}$ starts to increase upon cooling below around $T_2 = 60~{\rm K}$. The increase in the non-relaxing amplitude is indicative of a greater static component of the magnetism as temperatures are lowered, since, in the absence of dynamics, a muon whose spin lies parallel to the local field at its stopping site will not be depolarised.

The behaviour of the fast relaxation component is somewhat different; its amplitude displays an asymmetric peak, with a sudden increase upon cooling through $T_{\rm N}$ followed immediately by a gradual decrease as temperature is reduced further. The peak in the fast relaxation amplitude may indicate the freezing of dynamics accompanying the transition to magnetic LRO.

We now turn to the oscillations in the measured asymmetry within the temperature range $60\lesssim T \lesssim 135~{\rm K}$ [figure~\ref{SFOOraw}(b) inset]. In a quasistatic magnetic field, with magnitude $B$, a muon will undergo Larmor precession with frequency $\nu$ given by $2 \pi \nu = \gamma_{\mu} B$ where $\gamma_{\mu} = 2 \pi \times 135.5~{\rm MHz/T}$ is the muon gyromagnetic ratio. This intermediate temperature range is approximately bounded by the N\'{e}el temperature $T_\mathrm{N}$ and  secondary ordering temperature $T_2$ as reported in previous magnetic measurements [\onlinecite{paul13a}]. Data in this temperature range were fitted to the damped oscillatory relaxation function
\begin{equation}\label{eqn:osc}
A(t)=A_{\rm osc}e^{-\lambda_{\rm osc}t}\cos{(2 \pi \nu t + \varphi)} +A_{\rm rel}e^{-\lambda_{\rm rel}t}  +A_{\rm bg} ,
\end{equation}
for $t\leq 1~\mathrm{\mu s}$. For our detector geometry there should be zero phase offset, and indeed the data were fitted well with $\varphi$ set to zero (significantly, this was not the case for $B={\rm Y, In}$ as discussed below). The fitted oscillatory amplitudes were found not to vary significantly with temperature, and so were fixed to the average value $A_{\rm osc} = 3.2\%$. The total transverse component of the asymmetry data is therefore $A_{\bot} = A_{\rm lost} + A_{\rm osc} \approx 16\%$, which is approximately two-thirds of the high-temperature total asymmetry $A_2=23.3 \pm 0.3 \%$ (from equation \ref{eqn:FD}), as expected for a powder sample. The single precession frequency is ascribed to the presence of one class of muon stopping site which gives rise to a single oscillatory component in the asymmetry data (see section \ref{sec:site}). 

The fitted precession frequency $\nu$ drops steadily upon warming from 65~K towards $T_{\rm N}$, which coincides with the recovery of initial asymmetry as LRO is destroyed, in a typical fashion for an ordered system [figure~\ref{fig:SFOOfit}(d)]. The fitted frequency values are an effective order parameter, so for $T \geq 65~{\rm K}$ they were in turn fitted to the phenomenological expression
\begin{equation}\label{eqn:freq}
\nu (T) = \nu_0 \left[ 1 - \left( \frac{T}{T_{\rm N}} \right)^{\alpha~} \right] ^{\beta},
\end{equation}
yielding a N\'{e}el temperature $T_{\mathrm{N}}=135 \pm 2~\mathrm{K}$ and a zero-temperature frequency $\nu(0)=50 \pm 1~\mathrm{MHz}$ for $\alpha=4.2 \pm 0.9$ and $\beta = 0.42 \pm 0.07$, as shown in figure~\ref{fig:SFOOfit}(d). Here, the fitted value of $\beta$ is slightly larger than that expected for the critical exponent of a 3D Heisenberg antiferromagnet. Overdamped oscillations were also visible in asymmetry data for temperatures just below $T_2$, where the fitting function was able to capture an initial increase in asymmetry at very early times, but not to extract a reliable value of $\nu$. Oscillations cease to be resolvable altogether as temperatures are lowered further, indicating a broad range of internal magnetic fields experienced by the muon ensemble resulting in loss of coherent spin precession. This is evidence that supports the coexistence of the AF1 and AF2 phases proposed in reference~[\onlinecite{paul13b}]; if magnetic spin directions simply reorientated to adopt the AF2 configuration, then we would expect to continue to resolve oscillations in the $\mu$SR data, albeit with different precession frequencies.  

We may contrast these new $\mu$SR results to those of the related DP Sr$_2$CoOsO$_6$, reported in reference [\onlinecite{yan14}]. 
In that case, exchange interactions are predominantly intra-sublattice [\onlinecite{yan14,morrow13}] and in the intermediate temperature region $67 \leq T \leq 108~{\rm K}$ magnetic ordering is achieved by \textit{average} effective moments of dynamically fluctuating Co and Os spins, which precludes the appearance of coherent muon-spin precession. The Co moments adopt a complex noncollinear AFM spin arrangement below around 67~K. However, Os ions freeze into a glassy randomly canted state at around 6~K.
In SFOO, the presence of muon-spin precession oscillations in the AF1 phase is consistent with simultaneous quasistatic magnetic ordering on both the Fe and Os sublattices, and their absence below $T_2$ is further evidence for phase coexistence. We see no evidence of any further spin freezing in SFOO for temperatures down to 5~K.

\section*{\SYOO{} and \SIOO{}}

\begin{figure}[t]
\centering
\includegraphics[width=\linewidth]{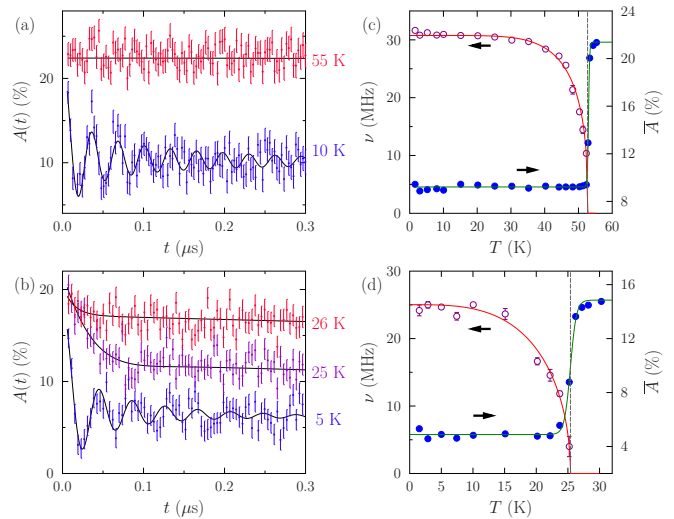}
\caption{\label{fig:combo} Example spectra for (a) SYOO and (b) SIOO at selected temperatures, showing oscillations for $T \lesssim 53$ and 25~K, respectively. (c) and (d) Temperature dependence of the fitted characteristic oscillatory frequency $\nu$, and the time-averaged asymmetry $\overline{A}$. Solid lines are fits described in the text and dashed lines indicate $T_{\rm N}$. }
\end{figure}

Example spectra for SYOO and SIOO measured at several temperatures are shown in figures~\ref{fig:combo}(a) and (b). Little change is observed in the asymmetry spectra at temperatures above $T_{\rm N}$ for each compound, where the data are well described by a Gaussian relaxation function (not shown). This is expected for the paramagnetic phase where rapidly fluctuating electronic dipole moments centred on the Os ions are motionally narrowed from the signal, and the muon response is dominated by  quasistatic magnetic fields due to the nuclear dipole moments. 

The temperature dependence of the time-averaged asymmetry $\overline{A} \equiv \langle A (t \leq 9.69~{\rm \mu s}) \rangle$ [figures~\ref{fig:combo}(c) and (d)] provides a means of determining the transition temperature. The broadened step function, equation~\ref{eqn:FD}, was again used to parametrise the drop of asymmetry upon cooling and the resultant fit yielded $T_{\rm N} = 53.0 \pm 0.2~{\rm K}$ for SYOO and $T_{\rm N} = 25.4 \pm 0.5~{\rm K}$ for SIOO, in good agreement with previous characterisation [\onlinecite{paul15}].

For temperatures below the transition at $T_{\rm N}$ magnetic moments order and oscillations become visible in the spectra, which indicates quasistatic LRO. The asymmetry spectra $A(t)$ within the ordered regime could be fitted to a function comprising a damped oscillatory component and a slowly relaxing exponential term equation~\ref{eqn:osc}. However, a non-zero phase $\varphi$ was found to be necessary to fit data for both SYOO and SIOO, with temperature-averaged values of $-47 \pm 2^{\circ}$ and $-44 \pm 4^{\circ}$, respectively. A cosinusoidal oscillation with phase $\varphi \approx -45^{\circ}$ closely resembles a zeroth order Bessel function of the first kind, which is indicative of incommensurate (IC) magnetic order [\onlinecite{amato97,dalmas06}]. Strictly speaking, the use of this function assumes
that the magnitude of the magnetic field strength at the muon site varies sinusoidally in space about an average of zero along with the wavelength determined by the magnetic wavevector being sufficiently large, relative to the crystallographic unit cell, that the implanted muons
thoroughly sample the magnetic order, particularly in the regions close to the nodes.

The fitting procedure was therefore repeated for both the Y and In compounds (for $t \leq 9.69~{\rm \mu s}$) with an exponentially damped Bessel function $J_0(2 \pi \nu t)$, plus a slowly relaxing exponential term:  
\begin{align}\label{eqn:bessel}
A(t)=A_\mathrm{osc}e^{-\lambda_\mathrm{osc}t} J_0 (2 \pi \nu t) +A_\mathrm{rel}e^{-\lambda_\mathrm{rel}t}  +A_\mathrm{bg} ,
\end{align}
where $\nu=(1 / 2\pi)\gamma_{\mu}B_{\rm max}$ now represents a characteristic precession frequency and $B_{\rm max}$ is the maximum local magnetic field magnitude. The oscillatory amplitude $A_{\rm osc}$ was fixed to the value of 11.6\% for SYOO, obtained via a global fit to the data measured for $T\leq 45~{\rm K}$, and left free to vary for SIOO [in both cases $A_{\rm osc} \approx (2/3) A(t=0)$, as expected for a powder sample]. Typical fits are presented in figures~\ref{fig:combo}(a) and (b) and the temperature dependence of the parameter $\nu$ is shown in figures~\ref{fig:combo}(c) and (d). The fitted initial asymmetry $A(t=0)$ is found to be temperature independent for both systems; no asymmetry is lost in these systems, unlike for SFOO. SYOO and SIOO only contain one magnetic species, so we expect internal magnetic field distributions to comprise smaller, more spatially uniform fields, and for the spin structures to promote less complicated spin dynamics relative to SFOO. Together these factors all engender smaller relaxation rates. 

The characteristic precession frequency $\nu$ acts as an effective order parameter for the LRO phase, and its temperature dependence was therefore fitted to the phenomenological expression, equation~\ref{eqn:freq}. For SYOO the resultant fit yields a value of the N\'{e}el temperature $T_{\rm N} = 52.7 \pm 0.1~{\rm K}$  (with $\alpha = 6.2 \pm 0.4$ and $\beta = 0.36 \pm 0.03$) and the fitted zero-temperature oscillatory frequency $\nu_0 = 30.8 \pm 0.1~{\rm MHz}$ equates to a maximum quasistatic magnetic field strength of $B_{\rm max}=0.227 \pm 0.001~{\rm T}$. For SIOO the paucity of data sets near the transition required the N\'{e}el temperature to be fixed to $T_{\rm N} = 25.4$~K, as obtained from the $\overline{A}$ fitting procedure. The fit yielded $\alpha = 2.4 \pm 0.6$, $\beta = 0.39 \pm 0.06$ and the zero-temperature oscillatory frequency $\nu_0 = 25.0 \pm 0.2 ~{\rm MHz}$, indicating a maximum quasistatic magnetic field strength of $B_{\rm max}=0.185 \pm 0.002~{\rm T}$ at the muon stopping sites. The fitted values of $\beta$ for these two systems are close to that expected for a 3D Heisenberg antiferromagnet.

\begin{figure*}[t]
\centering
\includegraphics[width=0.7\linewidth]{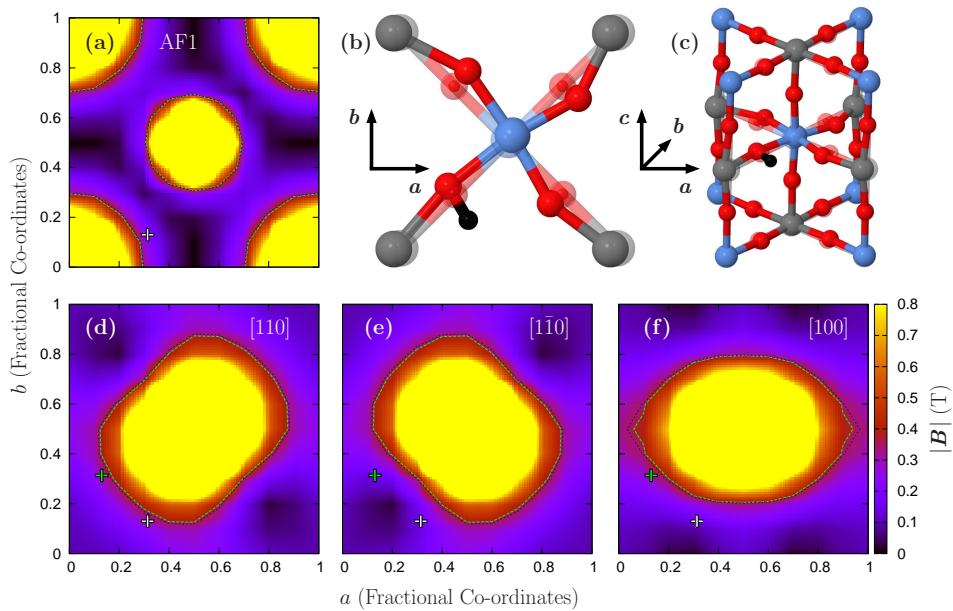}
\caption{\label{fig:heat} (a) Heat map showing the magnitude of the magnetic field $|{\bm B}|$ across the $ab$-plane (cross section taken at $c=1/2$) from dipole field simulations using the AF1 spin configuration within the undistorted unit cell, and magnetic moment magnitudes $\mu_{\rm eff}$ refined at $T=75~{\rm K}$ from reference~[\onlinecite{paul13b}]. The muon site determined by DFT calculations ($p1$) is shown as a white cross, and contours corresponding to experimentally observed magnetic field strengths $B_{\mu, {\rm exp}}$ are indicated (see main text). (b) and (c) chemical unit cell of Sr$_2$FeOsO$_6$ following insertion of the muon (original positions of ions are translucent). Data are visualised using Jmol [\onlinecite{Jmol}]. The cross-section in (b) is through the $c=1/2$ plane, containing the muon.  Blue represents Os ions, red O ions, grey Fe ions and black the muon (Sr ions are omitted for clarity). (d)--(f) Heat maps for SYOO, for various trial moment orientations within the $ab$-plane. The second class of muon site $p2$ is shown as a green cross. Plots for SIOO are very similar (not shown). }
\end{figure*}

Recently, IC magnetic order has been discovered in the related fcc DPs La$_2$NaRuO$_6$ ($4d^3$) and La$_2$NaOsO$_6$ ($5d^3$) on the basis of $\mu$SR and neutron scattering [\onlinecite{aczel13,aczel14}], where non-magnetic Na ions occupy the $B$ site, and $S=3/2$ (Ru/Os)$^{5+}$ ions occupy the $B'$ site. A suggested explanation for this is that a delicate balance between the relative signs and magnitudes of $J_{\rm nn}$ and $J_{\rm nnn}$ leads to the system lying on a boundary between AFM phases in the mean field theory $J_{\rm nn}$- $J_{\rm nnn}$ phase diagram, where only commensurate magnetic ordering is predicted [\onlinecite{lefmann01}].

The ratio of the fitted zero-temperature characteristic precession frequencies $\nu _0$  should be indicative of the relative sizes of the ordered moments of the $S=3/2$ Os$^{5+}$ ions for the two compounds. This ratio is given by
\begin{equation}
\frac{\nu_0(\rm Y)}{\nu_0(\rm In)}=\frac{30.8}{25.0}=1.23 \pm 0.01.
\end{equation}
In comparison, Rietveld refinements of neutron powder diffraction (NPD) data reveals ordered moment sizes of 1.91$\mu_{\rm B}$  and 1.77$\mu_{\rm B}$ for the Y and In compounds, respectively, yielding a ratio of $1.08 \pm 0.04$ [\onlinecite{paul15}]. These ratio values are in reasonable agreement; a possible cause of the difference between the two ratios is any extra hyperfine interaction between the muon and spin-polarised electron density in the Y compound. 
In addition, the difference in the unit cell dimensions between the two compounds [\onlinecite{paul15}] might lead to a slight relocation (in fractional coordinates) of the muon stopping site in each system.  In section \ref{sec:site} we identify candidate muon stopping sites for these compounds that occupy regions where the local magnetic dipole field strengths are varying rapidly. These factors could explain why muon-spin precession frequencies do not scale straightforwardly with moment size or unit cell volume.


\section{DFT and Dipole Simulations}
\label{sec:site}

Further information regarding the stopping site of implanted muons within the crystal unit cell plus the ensuing perturbation to its neighbouring environment allows a greater understanding of $\mu$SR results [\onlinecite{moller2013a}]. To this end, density functional theory (DFT) calculations have proved to be a valuable and powerful tool in the characterisation of muon states in host materials [\onlinecite{moller2013a,moller2013b,bonfa2015}].

In order to locate the muon  stopping site in SFOO, we made use of  the {\sc castep}  {\em ab  initio} package [\onlinecite{Clark-2005}], using ensemble DFT [\onlinecite{Payne-1992}] and on  the fly generated  Perdew-Becke-Ernerhof  functionals [\onlinecite{Perdew-1996}].  A $2\times3\times3$ Monkhorst-Pack  grid [\onlinecite{Monkhorst-1976}]  was used and the basis cut-off  energy was set to 489.8~eV.   Optimised lattice  parameters were found by relaxing the unit cell geometry without an implanted muon, yielding $c=7.810$~\AA{} and $a=b=5.501$~\AA{}, in close agreement with those values previously reported for this compound [\onlinecite{paul13a,paul13b}]. These parameters were fixed for subsequent calculations with a muon present. A muon was placed at  random  within  the  first  octant of  each unit  cell  subject  to  the constraint that it  not be located within 0.2~\AA$\,$  of an ion.  The unit cell geometry  and lattice  parameters were then optimised;  this was repeated 101 times with different starting positions so that the optimal location of the muon, i.e.\  the location with the lowest energy, could be found.  The optimal location of the  muon was found  to be  ${\bm r}_{\mu} =(0.314,0.130,0.499)$  in fractional co-ordinates, $1.003 \pm 0.001$~\AA{} from the nearest O ion.  The presence of the muon distorts the cell geometry, as can  be seen in figures~\ref{fig:heat}(b) and (c). In the following discussion we assume that all long range magnetic spin structures are robust enough that the presence of the muon and the subsequent lattice distortions do not perturb the local magnetic moments.

The consistency of the muon calculation performed in a single unit cell was checked by repeating the geometry  relaxation for the optimal location in a $2\times2\times1$  supercell.  The muon location was  found to be (0.308, 0.116,  0.499), which  is consistent with  the results  of the single cell (there is a 2\% change in the value of the $a$ co-ordinate and  a 10\%  change in  the value  of the  $b$ co-ordinate). Similar distortions of the Fe--O--Os bonds in the same $ab$ layer as the muon to those in  the single cell case  were observed. In addition, we  also found that  in the  $2\times2\times1$  supercell the  Fe--O--Os  bonds in  the neighbouring layer are  distorted to a similar degree  in the opposite direction. However, the magnetic ions are not significantly displaced and so this should not severely impact the dipole field at the candidate muon stopping site.

Having identified a candidate muon stopping site using DFT, dipole field simulations enable us to draw comparisons between theoretical and experimentally observed magnetic field strengths experienced by the muon ensemble. We are able to calculate the magnitude of the magnetic dipole field at a muon stopping site for a given spin structure using 
\begin{equation}
B^{\alpha}({\bm r})=\frac{\mu_0}{4\pi}\sum_{i,\beta}\frac{\mu_{{\rm eff},i}m_i^{\beta}}{R_i^3}
\left(\frac{3R_i^{\alpha} R_i^{\beta}  }{R_i^2}-\delta^{\alpha \beta}\right),
\end{equation}
where $\alpha$ and $\beta$ run  over $x$, $y$ and $z$ directions, $i$ labels a magnetic ion located at ${\bm r}_{i}$ within a Lorentz sphere centred at  ${\bm r}$, ${\bm R}_i={\bm  r}-{\bm  r}_i$ with  $R_i=|{\bm R}_i|$, $\mu_{{\rm eff},i}$  is the effective magnetic moment  of the ion  $i$ and $m_i^{\alpha}$ is the direction cosine of that moment along direction $\alpha$ [\onlinecite{Blundell-2009,yaouanc2011}]. The Lorentz radius was chosen to be 50$c$ since this produced results at the muon stopping site which converged to within $0.01~{\rm T}$ of those obtained using 100$c$ and 150$c$. For AFM spin configurations one does not need to consider additional contributions arising due to Lorentz or demagnetising fields, since there is no net magnetisation.

We calculate  $B_{\mu} \equiv |{\bm B}({\bm r}_{\mu})|$,  the magnitude  of the  magnetic field at the muon stopping site in SFOO, using spin orientations and moment sizes from reference~[\onlinecite{paul13b}], for  the AF1 phase with the $T=75$~K values   of   the   magnetic   moments  [$\mu ({\rm Fe})=1.83\mu_{\rm B}$, $\mu ({\rm Os})=0.48\mu_{\rm B}$]. Within the AF1 phase, moments residing on nn ions within the the $ab$-plane are aligned antiparallel to one another, and the spin sequence along the $c$-axis is $\uparrow \uparrow \uparrow \uparrow$.
This calculation yields $B_{\mu} = 0.21~{\rm T}$ for the undistorted unit cell, where no perturbations owing to the presence of the muon have been included, and $B_{\mu} = 0.15~{\rm T}$ for the distorted  crystal structure, where muon induced perturbations from the DFT relaxation procedure are included. 
The experimentally observed value of $B_{\mu}$ may be deduced using the fitting function, equation~\ref{eqn:freq}, which yields $\nu(T=75~{\rm K}) = 48.2 \pm 1.4~{\rm MHz}$, or equivalently $B_{\mu, {\rm exp}} = 0.36 \pm 0.01 ~{\rm T}$. Figure~\ref{fig:heat}(a) shows the simulated values of $|{\bm B}|$ within the $ab$-plane (for the $c=1/2$ cross section of the undistorted unit cell) containing the calculated muon stopping site, indicated by the white cross, plus the contours corresponding to the experimentally observed value $B_{\mu, {\rm exp}}$. Although the agreement between $B_{\mu, {\rm exp}}$ and $B_{\mu}$ shows a discrepancy, it is clear from figure~\ref{fig:heat}(a) that the proposed muon stopping site ${\bm r}_{\mu}$ lies in a region where $|{\bm B}({\bm r})|$ is rapidly changing with ${\bm r}$, and that the proposed muon site sits very close to the indicated contours corresponding to $B_{\mu, {\rm exp}} = 0.36 \pm 0.01~{\rm T}$ for the undistorted unit cell.

\begin{table*}[t]
\begin{tabular}{c|c|c|c}
Compound  &  \hspace{3mm} ${\bm m}$ \hspace{3mm}	&   Site  & \hspace{3mm}$|{\bm B}({\bm r}_{\mu})|$ (T)\hspace{3mm} \\ 
\hline 
Fe									& AF1		& $p1$  (undistorted) & 0.21  \\
$(B_{\mu, {\rm exp}}=0.35 \pm 0.01~{\rm T})$	& AF1		& $p1$  (distorted)   & 0.15    \\
\hline
				& [010]		& $p1$	& 0.29  \\
 				& [010]		& $p2$	& 0.13  \\
				& [100]		& $p1$ 	& 0.13  \\
Y				& [100]		& $p2$ 	& 0.29  \\
$(B_{\mu, {\rm exp}}=0.227 \pm 0.001~{\rm T})$	& [110]		& $p1$ 	& 0.28 \\
				& [110]		& $p2$ 	& 0.28 \\
				& [1$\bar{1}$0] & $p1$  & 0.15  \\
				& [1$\bar{1}$0] & $p2$  & 0.15  \\
\hline
 				& [010]		& $p$1	& 0.28  \\
				& [010]		& $p2$ 	& 0.12  \\
 				& [100]		& $p1$ 	& 0.12  \\
In 				& [100]		& $p2$ 	& 0.29  \\
$(B_{\mu, {\rm exp}}=0.185 \pm 0.002~{\rm T})$	& [110]		& $p1$ 	& 0.28 \\
				& [110]		& $p2$ 	& 0.28  \\
				& [1$\bar{1}$0] & $p1$ & 0.14  \\
				& [1$\bar{1}$0] & $p2$ & 0.14  \\
\end{tabular}
\caption{Table of simulated $B_{\mu}$ values for SFOO within the distorted and undistorted unit cell. The results of simulations are also provided for SYOO and SIOO, for various orientations of the magnetic moments ${\bm m}$ within the $ab$-plane, calculated at the two candidate muon stopping sites $p1$ and $p2$. Experimentally determined magnetic field strengths $B_{\rm \mu, exp}$ (see main text) are also provided for comparison. }
\label{tab:sims}
\end{table*}

We therefore conclude that the proposed muon site is plausible for SFOO. The discrepancy between the simulated $B_{\mu}$ and $B_{\mu, {\rm exp}}$ is larger for the simulated dipole fields within the distorted unit cell configuration. However, a potential pitfall of this calculation is that all unit cells within the Lorentz sphere will also be distorted, whereas in reality we expect the degree of distortion to decay rapidly with distance from the muon site. The long ranged nature of dipole fields means this could have a significant impact on the simulated value of $B_{\mu }$, particularly since ${\bm r}_{\mu}$ sits in a region of rapidly changing field strength, as discussed above. We also note that this approach assumes the magnetic field at the muon site is entirely dipolar, and neglects any hyperfine field coupling arising from spin-polarised electron density overlapping with the muon's wavefunction (this contribution is expected to be small for insulating systems such as these). Uncertainty in the moment sizes of each magnetic species adds further scope for discrepancy between simulated and observed magnetic field strengths.

We also calculate magnetic dipole fields within SYOO and SIOO using the AFM spin configuration previously determined via NPD measurements [\onlinecite{paul15}]. Magnetic moments are currently understood to assume a type I AFM arrangement, comprising alternating layers of ferromagnetically aligned spins, with adjacent layers along the crystallographic $c$-axis having opposite directions of spin polarisation. NPD data have constrained the directions of the Os magnetic moments to lie within the $ab$-plane, but have not resolved their exact orientations since the monoclinic distortion is small, and $a$ and $b$ are very nearly equal [\onlinecite{paul15}] (the same situation is encountered in the related compound Sr$_2$ScOsO$_6$ where the $B$ site is occupied by the $3d^0$ Sc$^{3+}$ ion [\onlinecite{paul15,taylor2015}]). 
In order to approximate candidate muon stopping sites in these compounds, which both crystallise in the monoclinic space group $P2_1 / n$, we consider two positions. These are $p1=(0.314,0.130,0.499)$ and $p2=(0.130,0.314,0.499)$, i.e.\ the fractional co-ordinates of ${\bm r}_{\mu}$ determined for SFOO, plus the position obtained by exchanging the $a$ and $b$ co-ordinates. For our simulations we used the type I AFM spin structure with moments orientated along various high symmetry directions within the $ab$-plane. Table~\ref{tab:sims} summarises our results, where effective magnetic moments and unit cell dimensions refined at low temperature were used for SYOO and SIOO [$\mu({\rm Os})=1.91\mu_{\rm B}$  and $\mu ({\rm Os})=1.77\mu_{\rm B}$, and $(5.78,5.81,8.18)~$\AA{} and  $(5.69, 5.70, 8.05)$~\AA{}, respectively], as reported in reference~[\onlinecite{paul15}].

For magnetic moment orientations along either the $a$ or $b$ axes (the [010] orientation is proposed in reference~[\onlinecite{paul15}]) our simulations predict the muon ensemble would experience two distinct magnitudes of local magnetic field $B_{\mu}$ [figure~\ref{fig:heat}(f) and Table~\ref{tab:sims}]. This situation would give rise to two resolvable oscillatory components in the muon data. However, as discussed in Sec.~\ref{sec:results}, the data are well described by a single oscillatory component. In contrast, $B_{\mu}$ at the sites $p1$ and $p2$ have essentially identical dipole field strengths when the moments are oriented along diagonal directions within the $ab$-plane [figures~\ref{fig:heat}(d) and (e)].

The $\mu$SR data lead us to consider the possibility of IC magnetic ordering within SYOO and SIOO. A muon ensemble residing at magnetically equivalent sites within a commensurate magnetic field texture would be expected to sample a field distribution comprising a narrow peak (as for the SFOO case). In contrast, a sinusoidally varying magnetic field strength (with IC wavevector) leads to muons sampling a magnetic field distribution [\onlinecite{amato97,major1986}] $p_{\rm IC}(B) \propto (B_{\rm max}^2 - B^2)^{-1/2}$ for fields $B < B_{\rm max}$ (and zero otherwise).  Since $p_{\rm IC}(B)$ is peaked towards $B_{\rm max}$, and at long times the Bessel function $J_0(\gamma_{\mu}B_{\rm max}t) \approx (1/ \sqrt{x})\cos (\gamma_{\mu}B_{\rm max}t- \pi / 4)$, comparison to the simulated $B_{\mu }$ values for commensurate ordering is physically justifiable.
The absolute agreement between simulated values of [110] and [1$\bar{1}$0] $B_{\mu}$ and the observed value $B_{\mu, {\rm exp}}$, calculated using the zero-temperature oscillatory frequency $\nu_0$ from the Bessel function fitting procedure ($0.227 \pm 0.001 ~{\rm T}$ and $0.185 \pm 0.002~{\rm T}$ for SYOO and SIOO, respectively), is reasonable. For instance, figure~\ref{fig:heat}(d) shows that the candidate muon sites lie in very close proximity to the contours corresponding to $B_{\mu, {\rm exp}}$ for the case where magnetic moments lie (anti-)parallel to the [110] or ${\bm a} + {\bm b}$ direction within SYOO. 

\section{Conclusion}
\label{sec:conc}

In conclusion, the results of our $\mu$SR study confirm the magnetic ordering temperature $T_{\rm N} = 135 \pm 2~{\rm K}$ for \SFOO{}, where the disappearance of oscillations in the muon asymmetry data for $T \leq T_2$ reveals a broader internal magnetic field distribution indicating coexistence of the antiferromagnetic phases AF1 and AF2, rather than a straightforward transition between the two phases. DFT calculations provide a candidate muon stopping site which dipole field simulations show to be consistent with our experimental results.

The N\'{e}el temperatures determined using $\mu$SR data for \SYOO{} and \SIOO{} are in excellent agreement with those previously obtained using molar magnetic susceptibility $\chi_{\rm m} (T)$ and specific heat capacity $C_{\rm p} (T)$ data [\onlinecite{paul15}]. However, within both the SYOO and SIOO systems our $\mu$SR measurements produce evidence for an incommensurate component to the magnetism, which was not observed in previous neutron powder diffraction experiments.

\section{Acknowledgements}

Part of this work was carried out at the Swiss muon source S$\mu$S, Paul Scherrer Institut, Villigen, Switzerland. We are grateful for the provision of beamtime, and to Alex Amato for muon experimental assistance. We thank the EPSRC (UK) and the John Templeton Foundation for financial support. We also wish to thank the Universities of Durham and York for use of HPC facilities.


\end{document}